\documentclass[12pt]{article}
\pdfoutput=1
\usepackage{epsf,amsfonts,amssymb,epsfig,amsmath,graphics,subfig}
\usepackage{hyperref,hep,graphicx, subfig}
\renewcommand{\text}[1]{#1}

\newcommand{\be}{\begin{equation}}
\newcommand{\ee}{\end{equation}}
\newcommand{\ben}{\begin{displaymath}}
\newcommand{\een}{\end{displaymath}}
\newcommand{\bea}{\begin{eqnarray}}
\newcommand{\eea}{\end{eqnarray}}
\newcommand{\ba}{\begin{align}}
\newcommand{\ea}{\end{align}}
\newcommand{\nn}{\nonumber \\}
\newcommand{\bi}{\begin{itemize}}
\newcommand{\ei}{\end{itemize}}

\begin{document}

\makeatletter
\renewcommand{\theequation}{\thesection.\arabic{equation}}
\@addtoreset{equation}{section}
\makeatother

\baselineskip 18pt

\begin{titlepage}

\vfill

\begin{flushright}
Imperial/TP/2011/JG/06\\
\end{flushright}

\vfill

\begin{center}
   \baselineskip=16pt
   {\Large\bf  Holographic helical superconductors}
  \vskip 1.5cm
  Aristomenis Donos$^1$ and Jerome P. Gauntlett$^2$\\
   \vskip .6cm
      \begin{small}
      \textit{Blackett Laboratory, 
        Imperial College\\ London, SW7 2AZ, U.K.\\Email: $^1$a.donos@imperial.ac.uk, $^2$j.gauntlett@imperial.ac.uk}
        \end{small}\\*[.6cm]

\end{center}

\vfill

\begin{center}
\textbf{Abstract}
\end{center}

\begin{quote}
We show that $D=5$ electrically charged $AdS$  black branes
can have instabilities associated with spatially modulated $p$-wave superconductors with helical structure. 
We show that the instabilities are present within $N=4$ $SU(2)\times U(1)$ gauged supergravity and hence within $D=10$ and $D=11$ supergravity.
\end{quote}

\vfill

\end{titlepage}
\setcounter{equation}{0}


\section{Introduction}
The AdS/CFT correspondence is a powerful framework for studying the properties
of strongly coupled quantum critical points, with potentially important applications
to condensed matter systems such as strongly correlated electrons or cold atoms. 
One focus of activity in such AdS/CMT studies has been the construction of
electrically charged $AdS$ black holes, describing CFTs at finite temperature and charge density,
that are holographically dual to superconducting states,
or more precisely to superfluid states. 

Black holes corresponding to $s$-, $p$- and $d$-wave superconductors, whose 
order parameters have angular momentum $l=0,1$ and $2$, respectively, have all been
constructed. In the $s$-wave superconducting black holes, the bulk charged fields dual to the order
parameter for the superconductivity, are scalar fields. They
have been constructed in phenomenological theories of gravity in \cite{Gubser:2008px,Hartnoll:2008vx,Hartnoll:2008kx}
and then in $D=10,11$ supergravity in 
\cite{Denef:2009tp,Gauntlett:2009dn,Gauntlett:2009bh,Gubser:2009qm}. $p$-wave
superconducting black holes have been constructed using either charged vector fields in the bulk  
\cite{Gubser:2008zu,Gubser:2008wv,Roberts:2008ns} or, alternatively, charged
two-forms \cite{Aprile:2010ge}, but not yet within $D=10/11$ supergravity\footnote{$p$-wave superconductors
have been discussed in the context of $D$-brane probes in \cite{Ammon:2008fc,Basu:2008bh,Peeters:2009sr}.}.
$p$-wave superconductivity is seen in some heavy fermion systems,
such as $UPt_3$, in $Sr_2RuO_4$ (e.g. see \cite{RevModPhys.75.657})
and in some organic materials, such as the Bechgaard-salt
$(TMTSF)_2PF_6$. $p$-wave superfluids are seen in $He_3$ and have also been recently observed
in Fermi-gases (e.g. see \cite{shin}). 
Holographic
$d$-wave superconducting black holes have been constructed in \cite{Benini:2010pr} using charged massive spin two-fields.
The embedding of the $d$-wave superconductors into $D=10,11$ supergravity is problematic because the consistency of interacting
massive spin two-fields requires an infinite number of bulk fields. The superconductivity in the high $T_c$ cuprates is well known to be 
of $d$-wave type.

All of these black hole solutions describe
spatially homogeneous superconducting states. However, it has long been known that it is possible to have superconducting states that are spatially inhomogeneous. Indeed in the Fulde-Ferrell-Larkin-Ovchinnikov (FFLO) phase, a Cooper pair consisting of two fermions with different Fermi momenta condenses leading to an order parameter with non-vanishing total momentum \cite{Fulde:1964zz,larkin:1964zz}. There are several systems in which it has been argued that the FFLO state is present but, it seems, not without some controversy.

A second focus of activity in AdS/CMT studies has been the construction of  
black holes that are dual to (non-superconducting) spatially modulated phases, which are also widely seen in condensed matter systems.
It has been shown that electrically charged black holes in $D=5$ can have instabilities corresponding to 
spatially modulated current density waves in $D=5$
in \cite{Nakamura:2009tf,Ooguri:2010kt,Ooguri:2010xs}
(for earlier related work see \cite{Domokos:2007kt}). Similar instabilities are also present in $D=4$ \cite{Donos:2011bh}
and in this case they are associated with  ``striped" black brane solutions which
are dual to phases with both current density waves and charged density waves, with the wave number for the latter being twice that
of the former. Examples of these instabilities were shown to
exist in D=10,11 supergravity in \cite{Donos:2011bh}, while analogous instabilities were observed in a probe-brane setting in
\cite{Bergman:2011rf}.
Most recently, it has been shown that magnetically
charged black branes
can also have spatially modulated instabilities \cite{Donos:2011qt}.

The purpose of this paper is to combine these two lines of development and show that spatially modulated 
superconducting states are possible within the context of AdS/CFT and moreover that they exist within string/M-theory.
The examples we discuss will be in the context of $D=5$ theories of gravity and correspond to spatially modulated $p$-wave superconductors
with a helical structure in the dual $d=4$ CFT.  
We will provide helical generalisations of both holographic  $p_x$-wave order \cite{Gubser:2008wv} and 
$(p_x+ip_y)$-wave order \cite{Gubser:2008zu}.
We will show that helical superconductors are possible using
two distinct mechanisms in $D=5$.

We first study each mechanism in simplified phenomenological models of gravity in $D=5$. 
In the first model, the bulk fields consist
of a metric, a $U(1)$ gauge field and the order parameter, spontaneously breaking the dual global $U(1)$ symmetry, is provided by a charged two-form satisfying a self-duality equation, 
similar to \cite{Aprile:2010ge}. In the second model
they consist of a metric, a $U(1)$ gauge-field and $SU(2)$ gauge-fields as in \cite{Zayas:2011dw}. For this model, generically the black branes
are charged with respect to $U(1)\times U(1)\subset U(1)\times SU(2)$ and the order parameter is provided by 
the charged $SU(2)$ gauge-fields, breaking $U(1)\times U(1)\to U(1)$. For one particular case, the background is just charged with respect to
the $U(1)$ factor and hence preserves the full $U(1)\times SU(2)$ global symmetry. For this case the 
instability preserves the $U(1)$ symmetry but breaks the $SU(2)$ symmetry, reminiscent of what is seen 
in spiral spin density waves.

We will also show that helical superconducting black holes appear
in Romans' $N=4^+$ $SU(2)\times U(1)$ gauged
supergravity theory \cite{Romans:1985ps}. This is significant because Romans' theory arises as
a consistent Kaluza-Klein (KK) truncation of type IIB supergravity on an $S^5$ \cite{Lu:1999bw},
thus capturing a sector of $N=4$ $d=4$ SYM, and also of $D=11$ supergravity on the general class of 
$M_6$ \cite{Lin:2004nb} (see also \cite{OColgain:2010ev})
associated with $N=2$ $d=4$ SCFTs \cite{Gauntlett:2007sm}.
Thus, by showing that the superconducting black holes are present in Romans' theory we will have
demonstrated the existence
of helical superconductors in $D=10,11$ supergravity. 
We will calculate the critical temperature at which the superconductivity involving the $SU(2)$ gauge-fields appears in Romans' theory.
Our numerical techniques are not stable enough to determine whether or not the superconductivity involving the
two-form also occurs in Roman's theory; if it does it will be at a much lower temperature.
On the other hand, we will show that Romans' theory also has a neutral instability of Gubser-Mitra type
\cite{Gubser:2000ec,Gubser:2000mm}, which appears at a higher temperature. This model thus has a rich structure and it will
be interesting to study it in more detail, generalising the analysis of \cite{Donos:2011ut}.

Our strategy is to analyse linearised perturbations about electrically charged $AdS$ black branes in $D=5$
with no charged hair. 
We will only consider black branes, such as the AdS-RN black branes, which at zero temperature approach an
$AdS_2\times\mathbb{R}^2$ solution in the near horizon limit. We first look for spatially modulated 
instabilities by studying perturbations that violate the $AdS_2$ BF bound. This provides a sufficient (but not necessary)
criteria for instabilities of the full black brane solutions.
For some representative examples, we then construct normalisable zero modes about the full black brane solutions 
and determine the critical temperatures at which the spatially modulated instabilities set in. 

\section{The charged two-form model}\label{twosec}
We consider a $D=5$ theory coupling a metric to a gauge field $A$ and a complex two-form $C$ with Lagrangian
\begin{align}\label{eq:lag2f}
\mathcal{L}=&(R+12)\ast 1-\frac{1}{2}\,\ast F\wedge F-\frac{1}{2}\ast C\wedge \bar{C}
-\frac{{i}}{2m}C\wedge \bar{H}\,,
\end{align}
where a bar denotes complex conjugation and the field strengths are 
\begin{align}\label{ders}
F=dA,\qquad
H=dC+{i} \frac{q}{\sqrt 3}\,A\wedge C\,.
\end{align}
The equations of motion are given by
\begin{align}\label{2feom}
R_{\mu\nu}&=-4g_{\mu\nu}+\frac{1}{2}\left(F_\mu{}^\rho F_{\nu\rho}-\frac{1}{6}g_{\mu\nu}F_{\rho\sigma}F^{\rho\sigma}\right)
+\frac{1}{2}\left(C_{(\mu}{}^\rho\bar C_{\nu)\rho}-\frac{1}{6}g_{\mu\nu}C_{\rho\sigma}\bar C^{\rho\sigma}\right)\nn
d*F&=-\frac{q}{2\sqrt{3}m} C\wedge C\nn
\ast H&={i} mC
\end{align}
When $q=m=1$, the black hole backgrounds and the
{\it linearised} perturbations that we consider for this model, are also relevant for Romans' theory of
$N=4^+$ $SU(2)\times U(1)$ gauged supergravity, as we discuss in section \ref{romans}.
In \cite{Aprile:2010ge} spatially homogeneous $p$-wave superconductivity was investigated with $m=1$ and variable $q$
in a similar context.
Here we investigate spatially modulated $p$-wave superconductivity for general $m,q$.

The equations of motion \eqref{2feom}
admit a unit radius $AdS_5$ solution with vanishing matter fields, that
is dual to the vacuum state of a $d=4$ CFT with a global
$U(1)$ symmetry. 
The gauge-field $A$ is dual to the conserved current associated with the global $U(1)$ symmetry and has dimension $\Delta=3$.
The equation of motion for the two-form $C$
describes three complex massive propagating degrees of freedom corresponding to 
self-dual tensor operators\footnote{When Romans' theory is considered
as a consistent truncation of type IIB supergravity, the $\Delta=3$ operator dual to the two form $C$ has been
identified in $N=4$ $d=4$ SYM as 
being $Tr\Phi F^+_{\mu\nu}$, where $\Phi$ is a complex scalar and $F^+$ is the self-dual part of the YM field strength \cite{Aprile:2010ge}.}
 of dimension 
$\Delta_\pm=2\pm m$ in the dual CFT. 

The equations of motion also admit the electrically charged AdS-RN black brane solution 
\begin{align}\label{eq:RN_sol}
ds^{2}=-f(r)dt^{2}+\frac{dr^{2}}{f(r)}+r^{2}\,\left(dx_{1}^{2}+dx_{2}^{2}+dx_{3}^{2} \right),\qquad
A=a(r)\,dt,
\end{align}
where 
\begin{align}\label{fandb}
f&=r^{2}-\frac{r_{+}^{4}}{r^{2}}+\frac{\mu^{2}}{3}\,\left(\frac{r_{+}^{4}}{r^{4}}-\frac{r_{+}^{2}}{r^{2}} \right),\qquad
a(r)=\mu\,\left(1-\frac{r^{2}_{+}}{r^{2}}\right)\,.
\end{align}
Here $\mu$ is a chemical potential for the global $U(1)$ symmetry in the dual CFT. 
The (outer) event horizon is located at $r=r_{+}$ and the Hawking temperature is given by
$T=(6r_+^2-\mu^{2})/6\pi r_+$.
At zero temperature, when ${\sqrt 6}r_{+}=\mu$, the near horizon limit is the $AdS_{2}\times \mathbb{R}^{3}$ solution
\begin{align}\label{eq:AdS2_limit}
ds^{2}=L^2\left(-\rho^{2}\,dt^{2}+\frac{d\rho^{2}}{\rho^{2}}\right)+dx_{1}^{2}+dx_{2}^{2}+dx_{3}^{2},\qquad
A=\frac{1}{\sqrt{6}}\,\rho\,dt,
\end{align}
with radius squared given by $L^2=1/12$ and we have rescaled the coordinates $t\rightarrow \frac{1}{12}t$, 
$x_{i}\rightarrow (\sqrt{6}/\mu)\,x_{i}$.

\subsection{Instabilities for $AdS_{2}\times \mathbb{R}^{3}$}\label{thatone}
Consider the perturbation of the two-form around the $AdS_{2}\times \mathbb{R}^{3}$ solution \eqref{eq:AdS2_limit} given by\begin{align}\label{eq:AdS2_2form_ansatz}
\delta C=dr\wedge\left(u_{1}\,dx_{1}-v_{1}\,dx_{2} \right)+{i}\, dt\wedge\left(u_{2}\,dx_{1}-v_{2}\,dx_{2} \right)+dx_{3}\wedge\left(u_{3}\,dx_{2} +v_{3}\,dx_{1}\right)\,,
\end{align}
where $u_{i}, v_i$ are complex functions of $t$, $\rho$ and $x_{3}$. It is straightforward to show that
this decouples from other perturbations at linearised order. We will consider the following two types of spatially modulated
modes. In the first we take
\begin{align}\label{anone}
u_{i}=d_{i}\cos\left(kx_{3}\right),\qquad
v_i=d_{i}\sin\left(kx_{3}\right)\,,
\end{align}
with $d_i(t,\rho)$ real while in the second we take
\begin{align}\label{antwo}
u_{i}=d_{i}\,e^{ikx_{3}},\qquad
v_i=id_{i}\,e^{ikx_{3}}\,,
\end{align}
again with $d_i(t,\rho)$ real. These two cases correspond to what we shall call holographic
helical $p_x$ and $(p_x+ip_y)$-wave superconductors, respectively,
for reasons we discuss in section \ref{npws} below.
In both cases we find that the equation of motion for the two-form associated with \eqref{eq:lag2f} 
leads to three equations. Two of these can be used to solve for 
$d_1$ and $d_2$ in terms of $d_3$:
\begin{align}
d_1=-\frac{1}{m^{2}+k^2}\left(\frac{im}{\rho^2}{\cal D}_t+{k}{\cal D}_\rho\right)d_3,\qquad
d_2=\frac{1}{m^{2}+k^2}\left(ik{\cal D}_t-m\rho^2{}{\cal D}_\rho\right)d_3\,,
\end{align}
and we find that $d_3$ satisfies the second order equation
\begin{equation}
\left({\cal D}^2-L^2(m^{2}+k^2)+\frac{kq}{3\sqrt{2}\,m}\right)d_{3}=0\,.
\end{equation}
In these equations ${\cal D}$ is the unit radius $AdS_2$ derivative given by
\begin{align}\label{2dcov}
{\cal D}_\mu\equiv \nabla_\mu+i\frac{q}{\sqrt 3}A_\mu\,.
\end{align}

We thus conclude that these modes have an effective (unit-radius) $AdS_{2}$ mass given by
\begin{equation}
M^{2}=L^{2}\,\left(k^{2}+m^{2} \right)-\frac{kq}{3\sqrt{2}m}-\frac{q^{2}}{18}\,.
\end{equation}
The lightest mode is for $k_{min}=\sqrt{2}q/m$ and has
\begin{equation}\label{twoformBF}
M^{2}_{min}=\frac{m^{2}}{12}-\frac{q^{2}}{18m^{2}}\,\left(3+m^{2}\right)\,.
\end{equation}
For $q^{2}>3m^2/2$ we see that this violates the $AdS_2$ BF bound ${M}^{2}\ge-1/4$ and the instability will necessarily break the translational invariance of the background solution \eqref{eq:AdS2_limit}. For the case of the Romans' theory with $m=q=1$
we have ${M}^{2}_{min}=-5/36$ which satisfies the BF bound.

\subsection{Zero modes for the AdS-RN black brane}\label{twozbb}
The instabilities in the $AdS_2\times\mathbb{R}^3$ region that we saw in the last subsection will show up as 
instabilities of the AdS-RN black brane solution \eqref{eq:RN_sol} at some finite temperature depending on $k$. More
generally, at the temperature at which any instability appears the linearised perturbations about the AdS-RN black hole 
will, generically, admit a static normalisable zero mode, corresponding to the existence of a new branch of solutions.
We now analyse such zero modes.

We again consider a perturbation of the two-form as in \eqref{eq:AdS2_2form_ansatz}. We also consider the two cases
\eqref{anone}, \eqref{antwo} where the $d_i$ are real, time independent functions, $d_i=d_i(r)$, satisfying appropriate boundary conditions
at the black hole event horizon at $r=r_+$ and at the $AdS_5$ boundary $r\to\infty$.
After substituting into the two-form equation of motion \eqref{2feom} we again find that we can solve for $d_1$ and $d_2$ 
\begin{align}
d_{1}&=\frac{1}{f\left(m^{2}r^{2}+k^{2}\right)}\,\left(\frac{q}{\sqrt{3}}mra\,d_{3}-kf\,d_{3}^{\prime} \right)\nn
d_{2}&=-\frac{1}{m^{2}r^{2}+k^{2}}\,\left(\frac{q}{\sqrt{3}}ka\,d_{3}+mrf\,d_{3}^{\prime} \right)
\end{align}
and that $d_3$ satisfies the second order ODE
\begin{align}\label{2ndode}
\left(\frac{mrf}{(m^2r^2+k^2)}d_3'\right)'+\left(\frac{mrq^2a^2}{3f(m^2r^2+k^2)}-\frac{2m^2rkqa}{\sqrt{3}(m^2r^2+k^2)^2}-\frac{m}{r}\right)d_3=0\,.
\end{align}
One can check that for the spatially homogeneous case, when $k=0$, this is compatible with equation (3.12) of \cite{Aprile:2010ge}.

At the black hole event horizon we impose 
the boundary conditions
\begin{align}\label{eq:bc1_2form}
d_{3}=d_{3}^{0}+{\cal O}\left(r-r_{+}\right)
\end{align}
and we use the linearity of the ODE to choose $d_{3}^{0}=1$. It is straightforward
to check that the two-form $C$ and its field strength $H$ are both well defined at $r=r_+$.
As $r\to\infty$ we deduce that the asymptotic behaviour is of the form
\begin{equation}\label{eq:bc2_2form}
d_{3}=w_{1}\,r+\cdots w_{2}\,r^{-1}+\cdots \,.
\end{equation}
We are only interested in modes in which the operator dual to the two-form $C$ spontaneously breaks the
symmetry and so we demand $w_{1}=0$.

\begin{figure}
\centering
\subfloat[$q=1.812$]{\includegraphics[width=7cm]{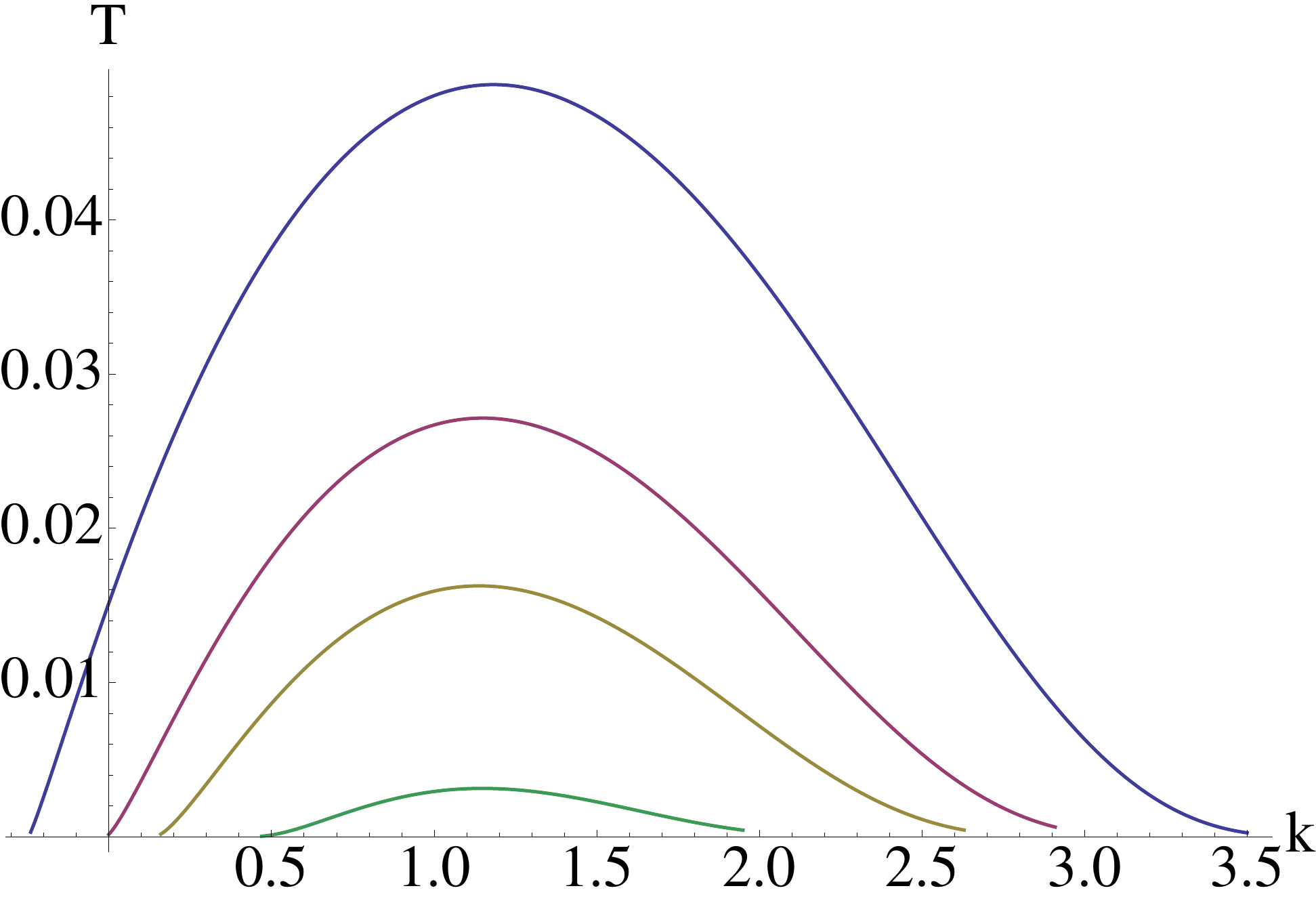}\label{fig:b1}}
\caption{Plots of critical temperatures $T$ versus $k$ for the existence of normalisable static perturbations of the two-form
about the AdS-RN black brane \eqref{eq:RN_sol} for $m=1$ and, from top to bottom, $q=2$, $q=1.812$, $q=1.7$ and $q=1.5$.
We have set $\mu=\sqrt{3}$.}\label{fig:b}
\end{figure}

We now return to the ODE \eqref{2ndode} with the boundary conditions \eqref{eq:bc1_2form}, \eqref{eq:bc2_2form}, which we have solved
using a shooting method. 
Since we have imposed two boundary conditions
on the second order ODE, for a given $k$ we expect zero mode solutions to exist, if at all,
at a specific temperature. 
We have made plots in Figure \ref{fig:b} of the 
temperature $T$ versus $k$ for which a normalisable mode exists for a representative range of the two-form charge $q$
and with $m=1$. The scale has been set by fixing $\mu=\sqrt{3}$.
The maximum of each curve indicate the critical temperature $T_c$ at which a spatially modulated
zero mode first appears. At $T=T_c$ a new branch of helical superconducting black branes will appear with
the spatial modulation, near $T=T_c$, set by $k_c$, the value of $k$ at $T_c$.

Observe that, for fixed $m$, as we increase $q$, we raise $T_c$. This mirrors the fact that
the corresponding modes in the $AdS_2\times\mathbb{R}^3$ region have bigger violations of the BF 
bound, as we see from \eqref{twoformBF}. 
Similarly fixing $q$ and increasing $m$ we find that $T_c$ decreases again mirroring
\eqref{twoformBF}. The case $q=m=1$ is interesting since these are the values relevant for 
Romans' theory. While we know that there are no violations of the BF bound in the 
$AdS_2\times\mathbb{R}^3$ region for this case , it is interesting to ask if there are zero modes in the full black hole solution.
We have not yet managed to stabilise our numerics for this case, but it is clear
from Figure \ref{fig:b} that if they do exist, they will certainly be at very low temperatures.

Another case worth commenting on further is $m=1$ and $q\approx 1.812$. 
For this case, we see from Figure \ref{fig:b} that the range of $k$
extends to include instabilities with $k=0$. Furthermore, for $m=1$ it is the smallest value of $q$ for which this happens.
In \cite{Aprile:2010ge} spatially homogeneous instabilities with $k=0$ were considered with $m=1$ and variable $q$
and it was argued that the value $q\approx 1.8$ is associated with a quantum critical 
point because for this value a $k=0$ mode is just becoming unstable at zero temperature.
However, in the larger set of perturbations with $k\neq 0$ that
we are considering here, we find that for this value of $q$ modes with $k>0$ become unstable at higher temperatures than the $k=0$ mode and hence
the value $q\approx 1.8$ is not singled out physically.

\subsection{The character of the $p$-wave superconductors}\label{npws}
We have established that there will be a new branch of superconducting black holes appearing at $T=T_c$ 
corresponding to
the linearised perturbation of $C$ given by \eqref{eq:AdS2_2form_ansatz} with $k=k_c$ for 
both \eqref{anone} and \eqref{antwo}.
Since $d_3$ is the only independent degree of freedom, the order parameter operator in the CFT dual to $C$ is governed by the
$d_3$ dependent terms in \eqref{eq:AdS2_2form_ansatz}. By taking the Hodge dual of the last term in \eqref{thatone} shows
that the operator is picking out a direction in the $x_1, x_2$ plane, corresponding to $p$-wave order, that 
rotates helically as one goes along the $x_3$ direction.  Let us now comment on the difference between the two cases
\eqref{anone} and \eqref{antwo}.

First consider the case \eqref{anone}. Recall that when $k=0$ this is an ansatz that was already considered
in \cite{Aprile:2010ge} and can be referred to as a holographic $p_x$-wave superconductor. In particular,
when $k=0$, after taking the Hodge-dual of the last term in \eqref{thatone} we see that that the operator dual to $C$ is picking out the 
(real) $dx_1$ direction. Furthermore, from \eqref{eq:lag2f} we can deduce that at second order the perturbation of $C$ will 
contribute to the energy momentum tensor, $T_{\mu\nu}$ ,with $T_{11}\ne T_{22}$ (and $T_{12}=0$) and hence the istotropy of the metric in the $x_1,x_2$ plane will
be broken, just as in the original holographic $p_x$-wave superconductor of \cite{Gubser:2008wv}. 
Allowing $k\ne0 $ in \eqref{anone}  it is clear that the dual operator picks out a direction
in the $x_1, x_2$ plane that rotates helically as we go along the $x_3$ direction. It is thus natural to call these holographic helical $p_x$-wave superconductors.
It is interesting to point out that when $k\ne 0$ we have $T_{12}\ne 0$, which we will provide an extra challenge in constructing fully back-reacted solutions.

We next consider the case \eqref{antwo}. When $k=0$ the dual operator is now picking out the complex $dx_1+idx_2$ direction 
and when $k\ne 0$ this is helically winding in the
$x_3$ direction. An important difference with the previous case is that the isotropy of the metric in the $x_1,x_2$ plane is now preserved
since $T_{11}=T_{22}$ and $T_{12}=0$ for all $k$. This isotropy is a feature of the original holographic ($p_x+ip_y$)-wave superconductors of \cite{Gubser:2008zu} and so
it is natural to call our $x_3$-dependent generalisations holographic helical ($p_x+ip_y$)-wave superconductors.

\section{The $SU(2)\times U(1)$ model}\label{onesec}
We now consider a $D=5$ theory coupling a metric, a $U(1)$ gauge field $B$ and
$SU(2)$ gauge-fields $A^\alpha$ with Lagrangian
\begin{align}\label{eq:lag2}
\mathcal{L}=&(R+12)\ast 1-\frac{1}{2}\ast G\wedge G-\frac{1}{2}\,\ast F^{\alpha}\wedge F^{\alpha}-\frac{\gamma}{2}F^{\alpha}
\wedge F^{\alpha}\wedge B\,, \end{align}
where the field strengths are 
\begin{align}\label{ders2}
G=dB,\qquad
F^{\alpha}=dA^{\alpha}-\frac{g}{\sqrt{2}}\epsilon_{\alpha\beta\gamma}\,A^{\beta}\wedge A^{\gamma}\,.
\end{align}
The equations of motion are given by
\begin{align}\label{eq:eomb2}
R_{\mu\nu}&=-4g_{\mu\nu}+\frac{1}{2}\left(G_\mu{}^\rho G_{\nu\rho}-\frac{1}{6}g_{\mu\nu}G_{\rho\sigma}G^{\rho\sigma}\right)
+\frac{1}{2}\left(F^{\alpha}_\mu{}^\rho F^\alpha_{\nu\rho}-\frac{1}{6}g_{\mu\nu}F^\alpha_{\rho\sigma}F^{\alpha\rho\sigma}\right)\nn
d\ast G&=-\frac{\gamma}{2}F^\alpha\wedge F^\alpha\nn
D\ast F^{\alpha}&=-\gamma\,F^{\alpha}\wedge G\,,
\end{align}
where $D\ast F^{\alpha}=d\ast F^{\alpha}+\sqrt{2}g\epsilon_{\alpha\beta\gamma}A^{\gamma}\wedge\ast F^{\beta}$.
The equations of motion admit a unit radius $AdS_5$ solution with vanishing gauge fields.
The corresponding dual $d=4$ CFT now has $SU(2)\times U(1)$ global symmetry and
$A^\alpha$ and $B$ are dual to the corresponding conserved currents, respectively, each with $\Delta=3$.
When $\gamma=g=1$, the black hole backgrounds and the
{\it linearised} perturbations that we consider for this model are also relevant for Romans' theory of
$N=4^+$ $SU(2)\times U(1)$ gauged supergravity, as we discuss in section \ref{romans}.
The general class of models \eqref{eq:lag2}
was recently studied in \cite{Zayas:2011dw} in the context of spatially homogenous
 $p$-wave superconductors, where the competition of
$p_x$-wave and $(p_x+ip_y)$-wave superconductivity was explored. We will comment on some of the results
of \cite{Zayas:2011dw} later.

The equations of motion admit the electrically charged $AdS$ black brane solutions
\begin{align}\label{eq:RN_solb2}
ds^{2}&=-f(r)dt^{2}+\frac{dr^{2}}{f(r)}+r^{2}\,\left(dx_{1}^{2}+dx_{2}^{2}+dx_{3}^{2} \right)\,,\nn
B&=b(r)=\mu_{1}\,\left(1-\frac{r_{+}^{2}}{r^{2}} \right)\,dt,\qquad
A^{3}=a(r)\,dt=\mu_{2}\,\left(1-\frac{r_{+}^{2}}{r^{2}} \right)\,dt, \notag\\
f&=r^{2}-\frac{r_{+}^{4}}{r^{2}}+\frac{1}{3}\left(\mu_{1}^{2}+\mu_{2}^{2}\right)\,\left(\frac{r_{+}^{4}}{r^{4}}-\frac{r_{+}^{2}}{r^{2}} \right)\,.
\end{align}
Note that we have allowed chemical potentials, $\mu_i$, for two $U(1)$ factors in the global $U(1)\times SU(2)$ symmetry group
of the dual CFT. Generically, the background \eqref{eq:RN_solb2}
breaks the global symmetry to $U(1)\times U(1)$ and the superconducting
instabilities that we consider will break this to a single $U(1)$.
As we explain in section \ref{romans} the solution with $\mu_{2}=\sqrt{2}\mu_{1}$ can be embedded in string or M-theory via Romans' theory. 

The temperature of these black holes is given by
$T=(6\,r_{+}^{2}-\mu_{1}^{2}-\mu_{2}^{2})/6\pi r_+$
with the extremal limit achieved for 
${\sqrt 6}r^{ext}_{+}=\sqrt{\mu_{1}^{2}+\mu_{2}^{2}}$. 
The near horizon limit of the zero temperature black hole is given by the $AdS_2\times\mathbb{R}^3$ solution
\begin{align}\label{eq:AdS2_limitb2}
ds_{5}^{2}=&L^2\left(-\rho^{2}\,dt^{2}+\frac{d\rho^{2}}{\rho^{2}}\right)+dx_{1}^{2}+dx_{2}^{2}+dx_{3}^{2}\,,\notag\\
A^{3}=&\frac{\sqrt{1-\chi^{2}}}{\sqrt{6}}\,r\,dt,\qquad B=\frac{\chi}{\sqrt{6}}\,\rho\,dt\,,
\end{align}
with $L^{2}=1/12$ and $0\leq\chi\leq 1$. 
Note that to obtain this we have defined $\chi=\mu_{1}/\sqrt{\mu_{1}^{2}+\mu_{2}^{2}}$ and rescaled $t\rightarrow \frac{1}{12}t$
and $x_i\to x_i/r^{ext}_+$.
We also note that for Romans' theory we should take $\chi=1/\sqrt{3}$.
In \cite{Zayas:2011dw} only black holes with $\chi=0$ (i.e. $\mu_1=0$) were studied.

\subsection{Instabilities for $AdS_{2}\times \mathbb{R}^{3}$}\label{thisone}
Consider the perturbation of the $SU(2)$ gauge-fields about the $AdS_{2}\times \mathbb{R}^{3}$ solution \eqref{eq:AdS2_limitb2}
of the form 
\begin{equation}\label{ofp}
\delta A^{1}+{i} \,\delta A^{2}=w_{1}\,dx_{1}+w_{2}\,dx_{2}\,,
\end{equation}
with $w_{i}$ being complex functions of $\left(t,\rho,x_{3}\right)$.
At the linearised level this perturbation decouples from other perturbations. 
We again consider two types of spatially modulated perturbations.

The first ansatz  is
\begin{align}\label{ofpx}
w_1=w \cos(kx_3),\qquad w_2=-w\sin(k x_3)\,,
\end{align}
with $w$ a real function of $x^\mu=(t,\rho)$. Notice that when $k=0$, we obtain a $D=5$ analogue of
the ansatz for the $p_x$-wave holographic superconductors of \cite{Gubser:2008wv} (used in 
\cite{Zayas:2011dw}), which at higher orders break the isotropy of  the metric in the $x_1, x_2$ plane. 
Switching on $k\ne 0$ gives rise to holographic helical $p_x$-wave order.
The second ansatz is 
\begin{align}\label{ofpxy}
w_1=w e^{ikx_3},\qquad w_2=iwe^{ik x_3}\,,
\end{align}
with $w$ a real function of $x^\mu=(t,\rho)$. When $k=0$, we obtain a $D=5$ analogue of the ansatz for the ($p_x+ip_y$)-wave 
holographic superconductors of \cite{Gubser:2008zu} (used in \cite{Zayas:2011dw}), which preserves a diagonal subgroup of the rotations in the  
$x_1,x_2$ plane and a $U(1)\subset SU(2)$. The corresponding metric preserves the
isotropy of the metric in the $x_1, x_2$ plane.  Switching on $k\ne 0$ gives rise to holographic helical ($p_x+ip_y$)-wave order.

After substituting into the equations of motion for the gauge-fields, \eqref{eq:eomb2}, we find that in both cases, at linearised order,
$w$ satisfies
\begin{align}
\left({\cal D}^2-L^2k^2-\frac{1}{\sqrt 6}\chi \gamma k\right)w=0\,,
\end{align}
where we are using the covariant derivative on the unit-radius $AdS_2$ space given by
\begin{align}\label{2dcov2}
{\cal D}_\mu\equiv \nabla_\mu-ig\sqrt{2}A^3_\mu\,.
\end{align}
We thus conclude that these modes have an
effective (unit radius) $AdS_{2}$ mass given by
\begin{equation}
M^{2}=L^{2}k^{2}+\frac{g^{2}}{3}\left(\chi^{2}-1\right)-\frac{1}{\sqrt{6}}\chi k\gamma\,.
\end{equation}
The minimum mass occurs at $k_{min}=\sqrt{6}\chi\gamma$ giving
\begin{equation}
M_{min}^{2}=\frac{g^{2}}{3}\left(\chi^{2}-1 \right)-\frac{1}{2}\chi^{2}\gamma^{2}\,.
\end{equation}
By suitable choice of backgrounds specified by $\chi$ and theories specified by $g,\gamma$ it is easy to violate
the BF bound $M^2\ge -1/4$. In Romans' theory, for example, 
with $\gamma=g=1$ and $\chi=1/\sqrt 3$ 
we have $M^2_{min}=-7/18$ which
violates the BF bound $M^2\ge -1/4$.

We next consider $\chi=1$. This case has $\mu_2=0$ and hence the background solution is only carrying electric
charge with respect
to the $U(1)$ factor and has vanishing $SU(2)$ fields. Nevertheless, provided that $\gamma>1/\sqrt 2$ we can still violate the
BF bound with spatially modulated instabilities. For this case the background preserves the $U(1)\times SU(2)$
global symmetry and the perturbation preserves the $U(1)$ but breaks the $SU(2)$. 

Another interesting case is
$\chi=0$ ($\mu_1=0$) in which the background is only electrically charged with respect to $U(1)\subset SU(2)$. In this case
the BF bound is violated provided that $g>\sqrt{3}/2$ and notice that these superconducting modes have $k_{min}=0$ and are not spatially
modulated. In fact these modes were studied in \cite{Zayas:2011dw} and we will return to them in section \ref{shortneut}.

\subsection{Zero modes for the AdS-RN black brane}\label{onezbb}
We now consider analogous perturbations of the one-forms about the black brane solutions given in
\eqref{eq:RN_solb2}. We again consider the spatially modulated superconducting perturbations as
in \eqref{ofp},\eqref{ofpx},\eqref{ofpxy} but now with $w$ a time independent, real function of $r$.
Substituting this ansatz into the $SU\left(2\right)$ gauge field equation of motion in
\eqref{eq:eomb2} leads to, at linearised order, the second order ODE
\begin{equation}\label{eq:c_equation}
\left(rfw^{\prime}\right)^{\prime}-\left(\frac{k^{2}}{r}-\frac{2g^{2}a^{2}r}{f}-\gamma k\, b^{\prime}\right)w=0\,.
\end{equation}
We now wish to solve this equation by imposing suitable boundary conditions. At the black hole 
event horizon, $r=r_+$, we have the expansion
\begin{equation}\label{eq:bc1}
w=w^{0}+{\cal O}\left(r-r_{+}\right)\,.
\end{equation}
Observe that the gauge-field is regular when $w^0\ne 0$. Furthermore, since the
equation is linear we can choose $w^0=1$. At $r\to \infty$, the asymptotic $AdS_5$ boundary,
we have the expansion
 \begin{equation}\label{eq:bc2}
w=g_0+\cdots+g_{1}{r^{-2}}+\cdots\,.
\end{equation}
Since we are interested only in the case in which the $SU(2)$ global symmetry current in the boundary
CFT, dual to $A^{\alpha}$, spontaneously develops an expectation value, we demand that $g_0=0$.

\begin{figure}
\centering
\includegraphics[width=8cm]{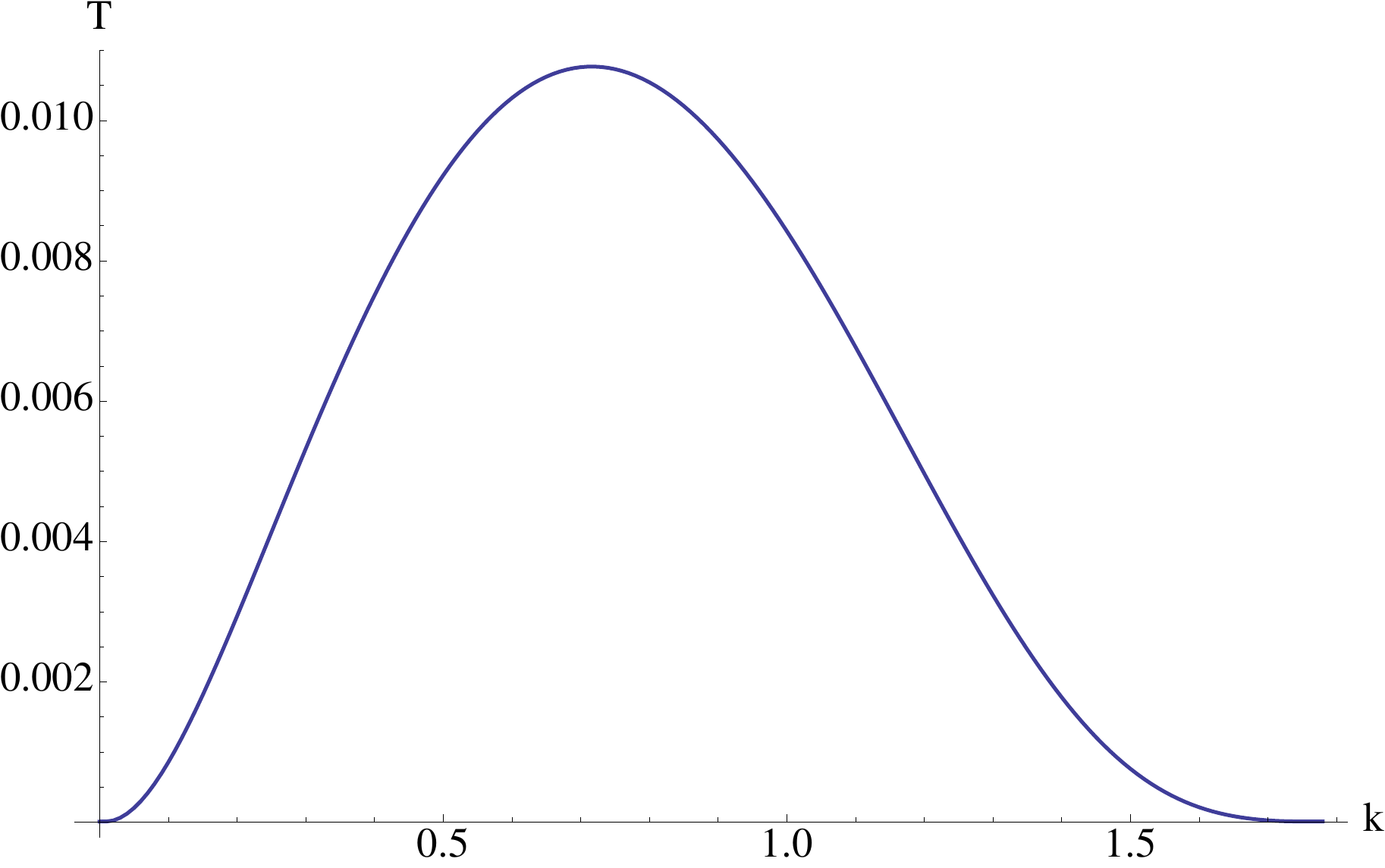}
\caption{Plot of critical temperatures $T$ versus $k$ for the existence of normalisable static perturbations of $A^1$, $A^2$
about the electrically charged black holes \eqref{eq:RN_solb2}. The plot is for Romans' theory with $\gamma=g=1$ and also 
$\mu_{2}=\sqrt{2}$ and $\mu_{1}=1$.}\label{fig:a}
\end{figure}

We have numerically solved equation \eqref{eq:c_equation} with these boundary conditions using a shooting method. 
The results are shown in Figure \ref{fig:a} for the Romans' case ($\gamma=g=1$, $\mu_{2}=\sqrt{2}\,\mu_{1}$ and we have scaled $\mu_1=1$) where we have plotted
the value of the temperature $T$ for which we can find such a function for any given momentum $k$. At $k=0$ the value of the temperature is very small, but non-zero. 
Observe that the highest critical temperature, $T_c\sim 0.011$ occurs for 
$k_c\sim 0.72$. At this temperature a new branch of black brane solutions will
exist that are dual to spatially modulated helical superfluid phases in the boundary CFT.

\subsection{Spatially modulated neutral instabilities}\label{shortneut}
Before concluding this section we would like to point out that in addition to the superconducting instabilities
that we have just discussed, the model \eqref{eq:lag2} also has neutral spatially modulated instabilities.
That this is the case can quickly be established as follows. Observe that we can consistently truncate $A^1=A^2=0$ and also set 
$A^3=\sqrt{2}B$
in the equations of motion for \eqref{eq:lag2}. After rescaling $\sqrt{3}B=A$, so that $A$ has a canonical kinetic term,  we find that the truncated 
model is exactly the same as that studied in \cite{Nakamura:2009tf} (after identifying $\gamma=2\sqrt{3}\alpha_{there}$).
In particular, for large enough values of $\gamma$ there will be spatially modulated neutral instabilities. 
In appendix \ref{neutral} we have presented a
few details of the analysis of these types of instabilities in the $AdS_2\times\mathbb{R}^3$ solution \eqref{eq:RN_solb2}, 
for general values of $\chi$.

The case of $\chi=0$ is relevant for the discussion of \cite{Zayas:2011dw}. In that paper an analysis of the competition
between $p_x$-wave and $(p_x+ip_y)$-wave superconductors was investigated. As we already pointed out above,
these superconductors are spatially homogeneous. 
(Note that to compare one
should identify our $g,\gamma$ with their $\alpha,\gamma$.)
Our analysis in appendix \ref{neutral} shows that
when $\chi=0$ and $\gamma>1$ the black holes will also have neutral spatially inhomogeneous instabilities.
It would be interesting to use these results 
to extend the analysis of \cite{Zayas:2011dw} to determine what are the true thermodynamically preferred
black holes.

\section{Romans' Theory}\label{romans}
Romans' $\mathcal{N}=4^{+}$ $SU(2)\times U(1)$ gauged supergravity \cite{Romans:1985ps} has bosonic fields 
consisting of a metric, gauge fields $A^{\alpha}$ and $B$, a complex two form $C$ and 
a neutral scalar field, $X$. Following the conventions of \cite{Gauntlett:2007sm} (and setting $m=1$), the Lagrangian is
\begin{align}\label{eq:lagrom}
\mathcal{L}=&R\ast 1-3X^{-2}\ast dX\wedge dX-\frac{1}{2}X^{4}\,\ast G\wedge G-\frac{1}{2} X^{-2}\,\left(\ast F^{\alpha}\wedge F^{\alpha}+\ast C\wedge \bar{C} \right)\notag\\
&-\frac{{i}}{2}C\wedge \bar{H}-\frac{1}{2}F^{\alpha}\wedge F^{\alpha}\wedge B+4\,\left(X^{2}+2X^{-1}\right)\,\ast 1\,,
\end{align}
where the field strengths are
\begin{align}\label{dersrom}
G=dB,\qquad
F^{\alpha}=dA^{\alpha}-\frac{1}{\sqrt{2}}\epsilon_{\alpha\beta\gamma}\,A^{\beta}\wedge A^{\gamma},\quad
H=dC+{i}\,B\wedge C\,.
\end{align}
Any solution of the equations of motion, 
which are explicitly given in \cite{Gauntlett:2007sm}, can be uplifted on an $S^5$ to obtain a solution of type IIB supergravity \cite{Lu:1999bw}.
They can also be uplifted on the general class of $M_6$ \cite{Lin:2004nb} (see also \cite{OColgain:2010ev}), 
corresponding to $N=2$ $d=4$ SCFTs, to obtain
an infinite class of solutions of $D=11$ supergravity \cite{Gauntlett:2007sm}.

As pointed out in \cite{Gauntlett:2007sm} there exists a consistent truncation to minimal five dimensional gauged supergravity after setting $X=1$, $C=A^1=A^2=0$ and $A^{3}=\sqrt{2}B$. Notice that the canonically normalised gauge field for minimal gauged supergravity
is then defined as $A=\sqrt{3}B$.
In particular, we notice that the electrically charged AdS-RN black brane solution \eqref{eq:RN_sol},\eqref{fandb} is a solution
of Romans' theory (with $A^3=\sqrt{2/3}a(r) dt$). Notice also that setting $q=1$ in \eqref{ders} agrees with \eqref{dersrom}.
Similarly setting $\mu_{2}=\sqrt{2}\mu_{1}$ in \eqref{eq:RN_solb2} we also obtain the same AdS-RN black hole solution of Romans' theory
after identifying $\mu=\sqrt{3}\mu_1$.

We can now consider perturbations about these AdS-RN black brane solutions. We find that 
the two-form perturbations and the one-form perturbations that we studied in the previous two sections remain
decoupled from other perturbations at {\it linearised} order also within Romans' theory\footnote{Going beyond linearised order for the two-form helical superconducting instability of section \ref{twosec}, one finds that the scalar
field $X$ is sourced, but not the $SU(2)$-gauge fields. For the instability of section \ref{onesec}, the scalar 
field is sourced, but not the charged two-form.}.
For the two-form perturbations we commented that, after setting $q=m=1$, if there are spatially modulated zero modes associated with the helical superconductors they would appear at very low temperatures.
On the other hand, for the one-form perturbations we saw that, after setting $\alpha=g=1$, the zero modes appear at $T_c\sim0.011$. 
Thus we conclude that helical superconducting black hole solutions can be found in Romans's theory and hence in string/M-theory.
In the next subsection we will show that Romans' theory has yet another instability, of a type first discussed by Gubser and Mitra 
\cite{Gubser:2000ec,Gubser:2000mm}, 
that happens at an even higher temperature than the helical superconducting instability. 

As first emphasised in \cite{Donos:2011ut}, identifying the highest critical temperature at which an instability sets in within a consistent
truncation is certainly
not sufficient to deduce the thermodynamically preferred phases of the dual CFTs. In principle one needs 
to construct all of the back reacted brane geometries and calculate
their free energies. These include the non-linear branches of black hole solutions
associated with the zero modes that we have found, but also possible branchings of these solutions. As in \cite{Donos:2011ut}
it is possible that the Gubser-Mitra type branch, for example, sprouts a superconducting branch at
lower temperatures. Furthermore, there could also be relevant
black holes associated with higher KK modes that are outside of the truncation to Romans theory.
We know for sure that at temperatures below the Gubser-Mitra critical temperature the dual CFTs cannot be described by the
AdS-RN black brane solutions (because they are unstable). This does {\it not} imply that 
black hole solutions associated with the superconducting zero modes for the AdS-RN black holes that we have constructed 
in
section \ref{onezbb} are physically irrelevant. While it does show these black holes do not
appear as thermodynamically preferred states arising from a second order phase transition,  it is still possible that 
they appear after a first order phase transition. It would be interesting
to know if this actually happens.

Before discussing the Gubser-Mitra instability we note that Romans' theory admits a one parameter
family of electrically charged $AdS_2\times \mathbb{R}^3$ solutions with
constant scalar field $X$, but unlike above, $X\ne 1$  \cite{Romans:1985ps} . 
It is plausible 
that these arise as the near horizon limits of zero temperature black brane solutions
which have not yet been constructed. In appendix \ref{line} we show that the one-form instabilities of section \ref{onezbb} are
present for all of these solutions.

\subsection{The Gubser-Mitra instability}
We now briefly discuss the zero modes associated with the Gubser-Mitra type instability of the AdS-RN black holes 
within Romans' theory. We consider the following 
perturbation around the black hole solution \eqref{eq:RN_solb2} with $\mu_2=\sqrt{2}\mu_1$:
\begin{align}
X=1+\frac{1}{\sqrt{6}}\,\delta\phi,\qquad
A^{3}=\sqrt{2}b(r)\,dt-\frac{1}{\sqrt{3}}\,\delta b\,dt,\qquad
B=b(r)\,dt+\sqrt{\frac{2}{3}}\delta b\,dt\,,
\end{align}
where $\delta\phi$ and $\delta b$ are functions of $r$.
The equations of motion at linearised order lead to the coupled differential equations
\begin{align}\label{eq:GM_eom}
\frac{1}{r^{3}}\,\left(r^{3}f\,\delta\phi^{\prime} \right)^{\prime}+2\,b^{\prime}{}^{2}\,\delta\phi+4\,\delta\phi+2\,b^{\prime}\,\delta b^{\prime}=0&\,,\nn
\left(2r^{3}b^{\prime}\,\delta\phi+r^{3}\,\delta b^{\prime}\right)^{\prime}=0\,.&
\end{align}
Near the horizon we impose the regular expansion
\begin{align}
\delta\phi=\phi_{+}^{(0)}+{\cal O}(r-r_+),\qquad
\delta b=b_{+}^{(1)}\,\left(r-r_{+}\right)+{\cal O}(r-r_+)^2
\end{align}
and we use the scaling symmetry of the linearised problem to set $\phi_{+}^{(0)}=1$.
The asymptotic behaviour as $r\to\infty$ is given by
\begin{align}
\delta\phi=\bar v_1\log(r)r^{-2}+\cdots +v_{1}r^{-2}+\cdots,\qquad
\delta b=g_0+\cdots+{g_{1}}{r^{-2}}+\cdots\,.
\end{align}
We are interested in the system spontaneously acquiring expectation values 
and so we demand that $\bar v_1=g_0=0$.
Using a shooting method, after setting $\mu_1=1$,
we find that the temperature for which such a solution to \eqref{eq:GM_eom} exists is at $T\approx 1/(2\pi)\approx 0.159$.

\section{Final Comments}
We have constructed linearised zero modes associated with spatially modulated black holes that are
dual to helical $p$-wave superconductors. We saw that the equations governing the $p_x$-wave and the $(p_x+ip_y)$-wave
cases were exactly the same at the linearised level. 
However, these cases will differ at higher orders and it will be interesting to see how they compete, generalising the investigations
of \cite{Gubser:2008wv,Roberts:2008ns,Zayas:2011dw}.
Extending the perturbative analysis beyond the linearised level will also illuminate additional features of
the helical superconductors, for example whether or not charge density waves are realised. The details will depend
on precisely which model one is considering; in Romans' theory, for example, the scalar field $X$ will become
activated. Of course, it would be most
interesting to go beyond a perturbative analysis and construct fully back reacted solutions.

More generally, this work combined with \cite{Nakamura:2009tf,Donos:2011bh,Donos:2011qt} indicates that
string/M-theory admits a very rich landscape of spatially modulated electrically and magnetically charged 
black hole solutions. This leads one to speculate that the generic ground states are not spatially homogeneous as hitherto thought.

\section*{Acknowledgements}
AD is supported by an EPSRC Postdoctoral Fellowship.
JPG is supported by an EPSRC Senior Fellowship and a Royal Society Wolfson Award. 
JPG would like to
thank the Aspen Center for Physics for hospitality and he acknowledges the 
support of the National Science Foundation Grant No. 1066293.

\appendix
\section{Neutral instabilities for the $SU(2)\times U(1)$ model}\label{neutral} 
For the model with $SU(2)\times U(1)$ symmetry, with Lagrangian given in \eqref{eq:lag2}, we consider the 
time independent perturbation
\begin{align}\label{neutansatz}
\delta A^3&=\frac{u(r)}{\sqrt{6}}\left[\cos(kx_3)\,dx_1-\sin(kx_3) \, dx_2\right]\notag\\
\delta B&=\frac{v(r)}{\sqrt{6}}\left[\cos(kx_3)\,dx_1-\sin(kx_3)\,dx_2 \right]\notag\\
\delta g_{tx_1}&=r\,h(r)\cos(kx_3)\notag\\
\delta g_{tx_2}&=-r\,h(r)\sin(kx_3)
\end{align}
around the one parameter family of $AdS_2\times \mathbb{R}^3$ solutions given in
\eqref{eq:AdS2_limitb2}. Such perturbations correspond to spatially modulated {\it non-superconducting}
instabilities generalising those studied in \cite{Nakamura:2009tf}.
At linearised order, the equations of motion give the coupled linear system of equations
\begin{align}\label{eq:linear_system}
12\,\left(r^2h^{\prime} \right)^{\prime}-k^2\,h+2r\,\left(\chi\,b^{\prime}+\sqrt{1-\chi^2}\,u^{\prime} \right)&=0\notag\\
12\,\left(r^2v^{\prime} \right)^{\prime}-k^2\,v+144\,\chi\,\left(rh\right)^{\prime}+2\sqrt{6}k\gamma \,\sqrt{1-\chi^2}\,u&=0\notag\\
12\,\left(r^2u^{\prime} \right)^{\prime}-k^2\,u+144\,\sqrt{1-\chi^2}\,\left(rh\right)^{\prime}+2\sqrt{6}k\,\gamma\,\sqrt{1-\chi^2}\,v+2\sqrt{6}k\,\gamma\,\chi\,u&=0
\end{align}

\begin{figure}
\centering
\includegraphics[width=8cm]{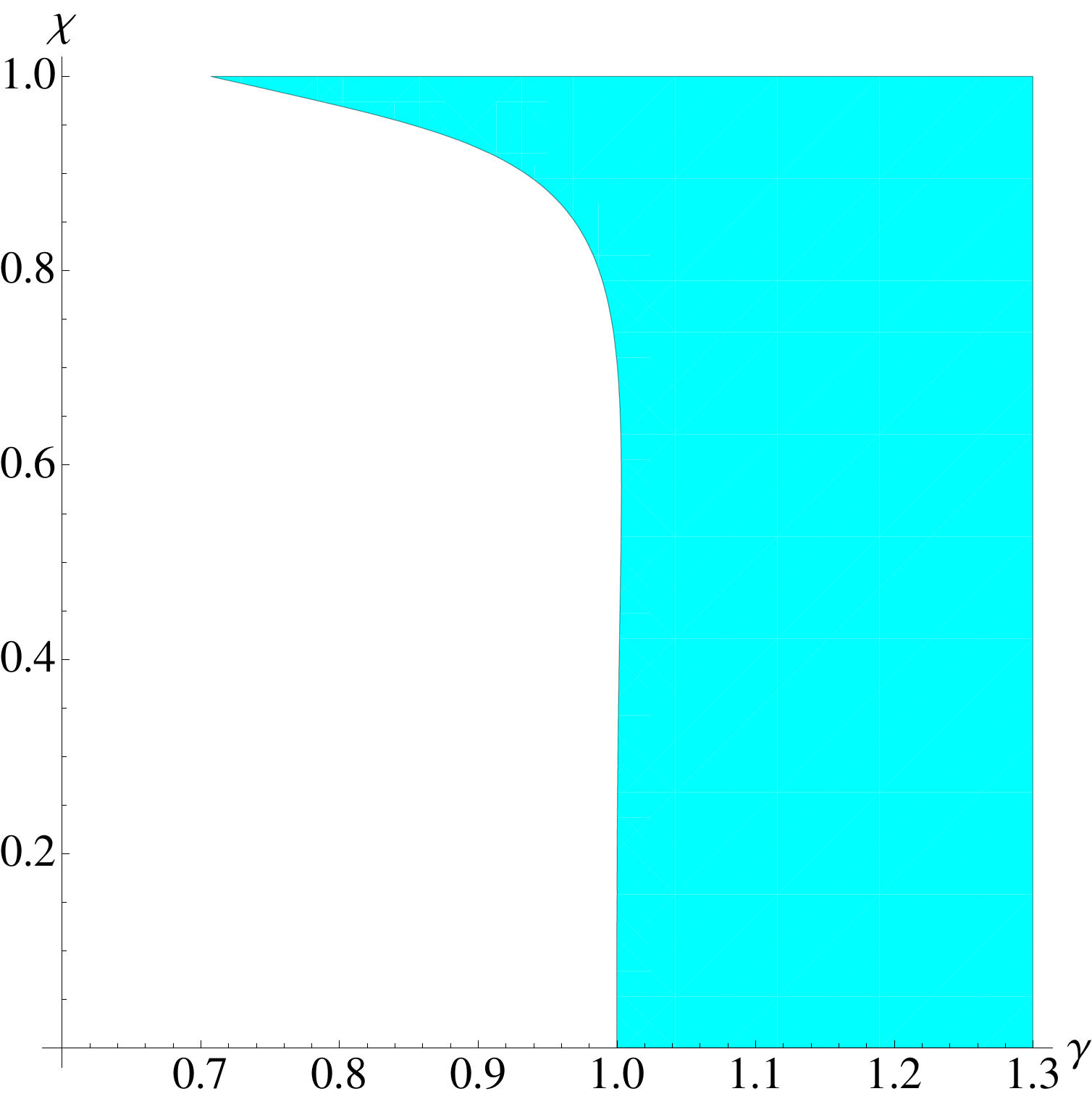}
\caption{The shaded region indicates values of $\left(\chi, \gamma\right)$ for which the $AdS_{2}\times \mathbb{R}^{3}$ solutions
\eqref{eq:AdS2_limitb2} have neutral instabilities of the form \eqref{neutansatz} which break translational and rotational invariance.}\label{fig:c}
\end{figure}

To determine the scaling dimensions of the dual operators 
we look for solutions of the form $\left(h,a,b \right)=\mathbf{v}\,r^{\lambda}$, with $\mathbf{v}$ a constant vector. Plugging this ansatz in the linear system \eqref{eq:linear_system} we obtain a matrix equation of the form $\mathbf{M} \mathbf{v}=0$, where $\mathbf{M}$ is a three by three matrix depending on $\lambda$, $\chi$, $\gamma$ and $k$. The possible values of $\lambda$ are then determined by the requirement that 
$\det\mathbf{M} =0$ has non-trivial solutions. 
If any $\lambda$ has a non-zero imaginary part then the $AdS_{2}\times\mathbb{R}^{3}$ background is unstable. 
In Figure \ref{fig:c} we have plotted a shaded region in the $\chi$ - $\gamma$ plane where these neutral instabilities exist.
In all cases, the instability is spatially modulated with $k\ne 0$.
Note that the $\chi=0$ case is relevant for the analysis of \cite{Zayas:2011dw}.

\section{Instabilities of a line of $AdS_{2}\times \mathbb{R}^{3}$ solutions}\label{line}
Romans' theory, with Lagrangian \eqref{eq:lagrom}, is known \cite{Romans:1985ps} to 
admit the following
one parameter family of $AdS_{2}\times \mathbb{R}^{3}$ solutions, with constant scalar 
field, $X=X_{0}>0$:
\begin{align}
ds^{2}_{5}&=L^{2}\,ds^{2}\left(-\rho^{2}\,dt^{2}+\frac{d\rho^{2}}{\rho^{2}}\right)+dx_{1}^{2}+dx_{2}^{2}+dx_{3}^{2}\notag\\
A^{3}&=X_{0}^{3/2}\frac{\sqrt{1+X_{0}^{3}}}{\sqrt{2}\left(2+X_{0}^{3}\right)}\,\rho\,dt,\quad B=\frac{1}{\sqrt{2}X_{0}^{3/2}\left(2+X_{0}^{3}\right)}\,\rho\,dt
\end{align}
where 
\begin{align}
L^{2}&=\frac{X_{0}}{4\left(2+X_{0}^{3}\right)}
\end{align}
Notice that when $X_0=1$ we recover the solution given in \eqref{eq:AdS2_limitb2} after setting $\mu_2=\sqrt{2}\mu_1$.
We now show that the one-form perturbations of the type discussed in section \ref{thisone}
violate the $AdS_2$ BF bound. On the other hand, the two-form perturbations of the type discussed in section \ref{thatone}
do not lead to such a violation for any value of $X_0$.

We again consider the one-form fluctuations as 
in \eqref{ofp},\eqref{ofpx},\eqref{ofpxy} with $w$ a real function of $t,\rho$.
Following a similar analysis as in section \eqref{eq:AdS2_limitb2} we find effective 
 (unit-radius) $AdS_{2}$ masses given by
\begin{equation}
M^{2}=k^{2}+X_{0}^{2}\left(-4+\frac{4}{2+X_{0}^{3}} \right)-\,\frac{2\sqrt{2}}{\sqrt{X_{0}}}k
\end{equation}
Notice that this always develops a minimum at
\begin{equation}
k_{min}=\sqrt{2X_{0}^{-1}}
\end{equation}
with
\begin{equation}
M_{min}^{2}=-\frac{2+3X_{0}^{3}+2X_{0}^{6}}{2\left(2+X_{0}^{3} \right)^{2}}<-\frac{1}{4}
\end{equation}
which is always violating the BF bound. It is worth noticing that this is a monotonically decreasing function of $X_{0}$ asymptoting to $-1$.

\bibliographystyle{utphys}
\bibliography{romans}{}

\providecommand{\href}[2]{#2}\begingroup\raggedright\begin{thebibliography}{10}

\bibitem{Gubser:2008px}
S.~S. Gubser, ``{Breaking an Abelian Gauge Symmetry Near a Black Hole
  Horizon},'' \href{http://dx.doi.org/10.1103/PhysRevD.78.065034}{{\em Phys.
  Rev.} {\bfseries D78} (2008) 065034},
\href{http://arxiv.org/abs/0801.2977}{{\ttfamily arXiv:0801.2977 [hep-th]}}.

\bibitem{Hartnoll:2008vx}
S.~A. Hartnoll, C.~P. Herzog, and G.~T. Horowitz, ``{Building a Holographic
  Superconductor},''
  \href{http://dx.doi.org/10.1103/PhysRevLett.101.031601}{{\em Phys. Rev.
  Lett.} {\bfseries 101} (2008) 031601},
\href{http://arxiv.org/abs/0803.3295}{{\ttfamily arXiv:0803.3295 [hep-th]}}.

\bibitem{Hartnoll:2008kx}
S.~A. Hartnoll, C.~P. Herzog, and G.~T. Horowitz, ``{Holographic
  Superconductors},''
  \href{http://dx.doi.org/10.1088/1126-6708/2008/12/015}{{\em JHEP} {\bfseries
  12} (2008) 015},
\href{http://arxiv.org/abs/0810.1563}{{\ttfamily arXiv:0810.1563 [hep-th]}}.

\bibitem{Denef:2009tp}
F.~Denef and S.~A. Hartnoll, ``{Landscape of superconducting membranes},''
  \href{http://dx.doi.org/10.1103/PhysRevD.79.126008}{{\em Phys. Rev.}
  {\bfseries D79} (2009) 126008},
\href{http://arxiv.org/abs/0901.1160}{{\ttfamily arXiv:0901.1160 [hep-th]}}.

\bibitem{Gauntlett:2009dn}
J.~P. Gauntlett, J.~Sonner, and T.~Wiseman, ``{Holographic superconductivity in
  M-Theory},'' \href{http://dx.doi.org/10.1103/PhysRevLett.103.151601}{{\em
  Phys. Rev. Lett.} {\bfseries 103} (2009) 151601},
\href{http://arxiv.org/abs/0907.3796}{{\ttfamily arXiv:0907.3796 [hep-th]}}.

\bibitem{Gauntlett:2009bh}
J.~P. Gauntlett, J.~Sonner, and T.~Wiseman, ``{Quantum Criticality and
  Holographic Superconductors in M- theory},''
  \href{http://dx.doi.org/10.1007/JHEP02(2010)060}{{\em JHEP} {\bfseries 02}
  (2010) 060},
\href{http://arxiv.org/abs/0912.0512}{{\ttfamily arXiv:0912.0512 [hep-th]}}.

\bibitem{Gubser:2009qm}
S.~S. Gubser, C.~P. Herzog, S.~S. Pufu, and T.~Tesileanu, ``{Superconductors
  from Superstrings},''
  \href{http://dx.doi.org/10.1103/PhysRevLett.103.141601}{{\em Phys. Rev.
  Lett.} {\bfseries 103} (2009) 141601},
\href{http://arxiv.org/abs/0907.3510}{{\ttfamily arXiv:0907.3510 [hep-th]}}.

\bibitem{Gubser:2008zu}
S.~S. Gubser, ``{Colorful horizons with charge in anti-de Sitter space},''
  \href{http://dx.doi.org/10.1103/PhysRevLett.101.191601}{{\em Phys. Rev.
  Lett.} {\bfseries 101} (2008) 191601},
\href{http://arxiv.org/abs/0803.3483}{{\ttfamily arXiv:0803.3483 [hep-th]}}.

\bibitem{Gubser:2008wv}
S.~S. Gubser and S.~S. Pufu, ``{The gravity dual of a p-wave superconductor},''
  \href{http://dx.doi.org/10.1088/1126-6708/2008/11/033}{{\em JHEP} {\bfseries
  11} (2008) 033},
\href{http://arxiv.org/abs/0805.2960}{{\ttfamily arXiv:0805.2960 [hep-th]}}.

\bibitem{Roberts:2008ns}
M.~M. Roberts and S.~A. Hartnoll, ``{Pseudogap and time reversal breaking in a
  holographic superconductor},''
  \href{http://dx.doi.org/10.1088/1126-6708/2008/08/035}{{\em JHEP} {\bfseries
  08} (2008) 035},
\href{http://arxiv.org/abs/0805.3898}{{\ttfamily arXiv:0805.3898 [hep-th]}}.

\bibitem{Aprile:2010ge}
F.~Aprile, D.~Rodriguez-G\'omez, and J.~G. Russo, ``{P-Wave Holographic
  Superconductors and Five-Dimensional Gauged Supergravity},''
  \href{http://dx.doi.org/10.1007/JHEP01(2011)056}{{\em JHEP} {\bfseries 01}
  (2011) 056},
\href{http://arxiv.org/abs/1011.2172}{{\ttfamily arXiv:1011.2172 [hep-th]}}.

\bibitem{Ammon:2008fc}
M.~Ammon, J.~Erdmenger, M.~Kaminski, and P.~Kerner, ``{Superconductivity from
  gauge/gravity duality with flavor},''
  \href{http://dx.doi.org/10.1016/j.physletb.2009.09.029}{{\em Phys. Lett.}
  {\bfseries B680} (2009) 516--520},
\href{http://arxiv.org/abs/0810.2316}{{\ttfamily arXiv:0810.2316 [hep-th]}}.

\bibitem{Basu:2008bh}
P.~Basu, J.~He, A.~Mukherjee, and H.-H. Shieh, ``{Superconductivity from D3/D7:
  Holographic Pion Superfluid},''
  \href{http://dx.doi.org/10.1088/1126-6708/2009/11/070}{{\em JHEP} {\bfseries
  11} (2009) 070},
\href{http://arxiv.org/abs/0810.3970}{{\ttfamily arXiv:0810.3970 [hep-th]}}.

\bibitem{Peeters:2009sr}
K.~Peeters, J.~Powell, and M.~Zamaklar, ``{Exploring colourful holographic
  superconductors},''
  \href{http://dx.doi.org/10.1088/1126-6708/2009/09/101}{{\em JHEP} {\bfseries
  09} (2009) 101},
\href{http://arxiv.org/abs/0907.1508}{{\ttfamily arXiv:0907.1508 [hep-th]}}.

\bibitem{RevModPhys.75.657}
A.~P. Mackenzie and Y.~Maeno, ``The superconductivity of
  ${\mathrm{sr}}_{2}{\mathrm{ruo}}_{4}$ and the physics of spin-triplet
  pairing,'' \href{http://dx.doi.org/10.1103/RevModPhys.75.657}{{\em Rev. Mod.
  Phys.} {\bfseries 75} (May, 2003) 657--712}.
  \url{http://link.aps.org/doi/10.1103/RevModPhys.75.657}.

\bibitem{shin}
Y.~Shin, C.~H. Schunck, A.~Schirotzek, and W.~Ketterle, ``{Phase diagram of a
  two-component Fermi gas with resonant interactions},'' {\em Nature}
  {\bfseries 451} (2008) 689.

\bibitem{Benini:2010pr}
F.~Benini, C.~P. Herzog, R.~Rahman, and A.~Yarom, ``{Gauge gravity duality for
  d-wave superconductors: prospects and challenges},''
  \href{http://dx.doi.org/10.1007/JHEP11(2010)137}{{\em JHEP} {\bfseries 11}
  (2010) 137},
\href{http://arxiv.org/abs/1007.1981}{{\ttfamily arXiv:1007.1981 [hep-th]}}.

\bibitem{Fulde:1964zz}
P.~Fulde and R.~A. Ferrell, ``{Superconductivity in a Strong Spin-Exchange
  Field},''
\href{http://dx.doi.org/10.1103/PhysRev1.35.A550}{{\em Phys. Rev.} {\bfseries
  135} (1964) A550--563}.

\bibitem{larkin:1964zz}
A.~I. Larkin and Y.~N. Ovchinnikov, ``{Nonuniform state of superconductors},''
{\em Zh. Eksp. Teor. Fiz.} {\bfseries 47} (1964) 1136--1146.

\bibitem{Nakamura:2009tf}
S.~Nakamura, H.~Ooguri, and C.-S. Park, ``{Gravity Dual of Spatially Modulated
  Phase},'' \href{http://dx.doi.org/10.1103/PhysRevD.81.044018}{{\em Phys.
  Rev.} {\bfseries D81} (2010) 044018},
\href{http://arxiv.org/abs/0911.0679}{{\ttfamily arXiv:0911.0679 [hep-th]}}.

\bibitem{Ooguri:2010kt}
H.~Ooguri and C.-S. Park, ``{Holographic End-Point of Spatially Modulated Phase
  Transition},'' \href{http://dx.doi.org/10.1103/PhysRevD.82.126001}{{\em Phys.
  Rev.} {\bfseries D82} (2010) 126001},
\href{http://arxiv.org/abs/1007.3737}{{\ttfamily arXiv:1007.3737 [hep-th]}}.

\bibitem{Ooguri:2010xs}
H.~Ooguri and C.-S. Park, ``{Spatially Modulated Phase in Holographic
  Quark-Gluon Plasma},''
  \href{http://dx.doi.org/10.1103/PhysRevLett.106.061601}{{\em Phys. Rev.
  Lett.} {\bfseries 106} (2011) 061601},
\href{http://arxiv.org/abs/1011.4144}{{\ttfamily arXiv:1011.4144 [hep-th]}}.

\bibitem{Domokos:2007kt}
S.~K. Domokos and J.~A. Harvey, ``{Baryon number-induced Chern-Simons couplings
  of vector and axial-vector mesons in holographic QCD},''
  \href{http://dx.doi.org/10.1103/PhysRevLett.99.141602}{{\em Phys. Rev. Lett.}
  {\bfseries 99} (2007) 141602},
\href{http://arxiv.org/abs/0704.1604}{{\ttfamily arXiv:0704.1604 [hep-ph]}}.

\bibitem{Donos:2011bh}
A.~Donos and J.~P. Gauntlett, ``{Holographic striped phases},''
  \href{http://dx.doi.org/10.1007/JHEP08(2011)140}{{\em JHEP} {\bfseries 08}
  (2011) 140},
\href{http://arxiv.org/abs/1106.2004}{{\ttfamily arXiv:1106.2004 [hep-th]}}.

\bibitem{Bergman:2011rf}
O.~Bergman, N.~Jokela, G.~Lifschytz, and M.~Lippert, ``{Striped instability of
  a holographic Fermi-like liquid},''
\href{http://arxiv.org/abs/1106.3883}{{\ttfamily arXiv:1106.3883 [hep-th]}}.

\bibitem{Donos:2011qt}
A.~Donos, J.~P. Gauntlett, and C.~Pantelidou, ``{Spatially modulated
  instabilities of magnetic black branes},''
\href{http://arxiv.org/abs/1109.0471}{{\ttfamily arXiv:1109.0471 [hep-th]}}.

\bibitem{Zayas:2011dw}
L.~A.~P. Zayas and D.~Reichmann, ``{A Holographic Chiral $p_x + ip_y$
  Superconductor},''
\href{http://arxiv.org/abs/1108.4022}{{\ttfamily arXiv:1108.4022 [hep-th]}}.

\bibitem{Romans:1985ps}
L.~J. Romans, ``{Gauged N=4 supergravities in five-dimensions and their
  magnetovac backgrounds},''
\href{http://dx.doi.org/10.1016/0550-3213(86)90398-6}{{\em Nucl. Phys.}
  {\bfseries B267} (1986) 433}.

\bibitem{Lu:1999bw}
H.~Lu, C.~N. Pope, and T.~A. Tran, ``{Five-dimensional N = 4, SU(2) x U(1)
  gauged supergravity from type IIB},''
  \href{http://dx.doi.org/10.1016/S0370-2693(00)00073-3}{{\em Phys. Lett.}
  {\bfseries B475} (2000) 261--268},
\href{http://arxiv.org/abs/hep-th/9909203}{{\ttfamily arXiv:hep-th/9909203}}.

\bibitem{Lin:2004nb}
H.~Lin, O.~Lunin, and J.~M. Maldacena, ``{Bubbling AdS space and 1/2 BPS
  geometries},'' \href{http://dx.doi.org/10.1088/1126-6708/2004/10/025}{{\em
  JHEP} {\bfseries 10} (2004) 025},
\href{http://arxiv.org/abs/hep-th/0409174}{{\ttfamily arXiv:hep-th/0409174}}.

\bibitem{OColgain:2010ev}
E.~O~Colgain, J.-B. Wu, and H.~Yavartanoo, ``{On the generality of the LLM
  geometries in M-theory},''
  \href{http://dx.doi.org/10.1007/JHEP04(2011)002}{{\em JHEP} {\bfseries 04}
  (2011) 002},
\href{http://arxiv.org/abs/1010.5982}{{\ttfamily arXiv:1010.5982 [hep-th]}}.

\bibitem{Gauntlett:2007sm}
J.~P. Gauntlett and O.~Varela, ``{D=5 $SU(2)$xU(1) Gauged Supergravity from
  D=11 Supergravity},''
  \href{http://dx.doi.org/10.1088/1126-6708/2008/02/083}{{\em JHEP} {\bfseries
  02} (2008) 083},
\href{http://arxiv.org/abs/0712.3560}{{\ttfamily arXiv:0712.3560 [hep-th]}}.

\bibitem{Gubser:2000ec}
S.~S. Gubser and I.~Mitra, ``{Instability of charged black holes in anti-de
  Sitter space},''
\href{http://arxiv.org/abs/hep-th/0009126}{{\ttfamily arXiv:hep-th/0009126}}.

\bibitem{Gubser:2000mm}
S.~S. Gubser and I.~Mitra, ``{The evolution of unstable black holes in anti-de
  Sitter space},'' {\em JHEP} {\bfseries 08} (2001) 018,
\href{http://arxiv.org/abs/hep-th/0011127}{{\ttfamily arXiv:hep-th/0011127}}.

\bibitem{Donos:2011ut}
A.~Donos and J.~P. Gauntlett, ``{Superfluid black branes in $AdS_4\times
  S^7$},'' \href{http://dx.doi.org/10.1007/JHEP06(2011)053}{{\em JHEP}
  {\bfseries 06} (2011) 053},
\href{http://arxiv.org/abs/1104.4478}{{\ttfamily arXiv:1104.4478 [hep-th]}}.

\end{thebibliography}\endgroup
\end{document}